\pdfoutput=1

\documentclass[3p,times]{elsarticle}

\usepackage{ecrc}

\usepackage[breaklinks]{hyperref}
\usepackage[utf8x]{inputenc}
\usepackage{subfig}
\usepackage{amsmath}
\usepackage{graphicx}

\volume{00}

\firstpage{1}

\journalname{Nuclear Physics A}

\runauth{T. Lappi and H. Mäntysaari}

\jid{nupha}

\usepackage{amssymb}

\biboptions{comma,square,sort&compress}

\usepackage[figuresright]{rotating}

\newcommand{\xt}{{\mathbf{x}_T}}

\newcommand{\bt}{{\mathbf{b}_T}}

\newcommand{\xtp}{{\mathbf{x}'_T}}
\newcommand{\btp}{{\mathbf{b}'_T}}

\newcommand{\ztp}{{\mathbf{z}'_T}}

\newcommand{\zt}{{\mathbf{z}_T}}

\newcommand{\qt}{{\mathbf{q}_T}}
\newcommand{\kt}{{\mathbf{k}_T}}

\newcommand{\utt}{{u_T}} 

\newcommand{\ud}{\, \mathrm{d}}

\newcommand{\nc}{{N_\mathrm{c}}}

\newcommand{\cf}{C_\mathrm{F}}

\newcommand{\nr}[1]{(\ref{#1})}

\newcommand{\gev}{\ \textrm{GeV}}

\newcommand{\qs}{Q_\mathrm{s}}
\newcommand{\qso}{Q_\mathrm{s0}}

\newcommand{\lqcd}{\Lambda_{\mathrm{QCD}}}
\newcommand{\as}{\alpha_{\mathrm{s}}}

\newcommand{\fig}{Fig.~}

\newcommand{\eq}{Eq.~}

\newcommand{\re}{Ref.~}

\begin{document}
\hypersetup{pdfauthor={T. Lappi and H. M\"antysaari},pdftitle={Forward dihadron correlations in the Gaussian approximation of JIMWLK}}
\begin{frontmatter}



\dochead{}

\title{Forward dihadron correlations in the Gaussian approximation of JIMWLK}


\author{T. Lappi}
\address{
Department of Physics, %
 P.O. Box 35, 40014 University of Jyv\"askyl\"a, Finland and \\
 Helsinki Institute of Physics, P.O. Box 64, 00014 University of Helsinki,
Finland
}

\author{H. M\"antysaari}
\address{
Department of Physics, %
 P.O. Box 35, 40014 University of Jyv\"askyl\"a, Finland
}



\begin{abstract}
We compute forward dihadron azimuthal correlations
in deuteron-gold collisions
 using a Gaussian approximation
for the quadrupole operator. The double parton scattering contribution is found 
to be part of our dihadron calculation. We obtain a
 good description of the PHENIX data for the azimuthal-angle 
dependent away side peak and a relatively good estimate for the 
pedestal contribution.
\end{abstract}

\begin{keyword}
Dihadron correlations, JIMWLK, BK


\end{keyword}

\end{frontmatter}


\section{Introduction}
\label{introduction}

At high energy or, equivalently, small~$x$, the interactions
of hadrons are expected to be dominated by  nonlinear strong 
color fields. A convenient effective theory approach to studying
these color fields is provided by the Color Glass Condensate
(for a review see e.g.~\cite{Lappi:2010ek}).



Measurements of correlations between two forward 
hadrons in dAu collisions at 
RHIC~\cite{Braidot:2011zj,Adare:2011sc}
seem to show indications of ``initial state'' 
or ``cold nuclear matter'' effects that are significantly stronger than
in pp collisions or at central 
rapidities.
The upcoming LHC proton-lead collisions will 
provide more opportunities to study these phenomena in a wider
kinematical range.

These observations have provided an impetus for renewed interest
in the gluonic correlations included in the JIMWLK
evolution, see e.g.~\cite{Dumitru:2011vk}.
In particular it was argued
that the result of a full JIMWLK evolution, at finite $\nc$,
can quite accurately  be captured by the so called Gaussian 
approximation, relating higher point Wilson line correlators
to the two-point function.
These recent theoretical developments were not fully reflected
in the pioneering calculations of dihadron correlations 
in~\cite{Marquet:2007vb,Albacete:2010pg} 
where a factorized approximation is derived
in a certain kinematical limit. 
The main purpose of this work (see \re \cite{Lappi:2012nh}) is
to implement the Gaussian approximation, which so far has only been tested 
for particular coordinate space configurations, in a full calculation 
of the dihadron correlation.

\section{Single inclusive baseline}

\begin{figure}
\begin{minipage}[t]{0.48\linewidth}
\centering
\includegraphics[width=1.05\textwidth]{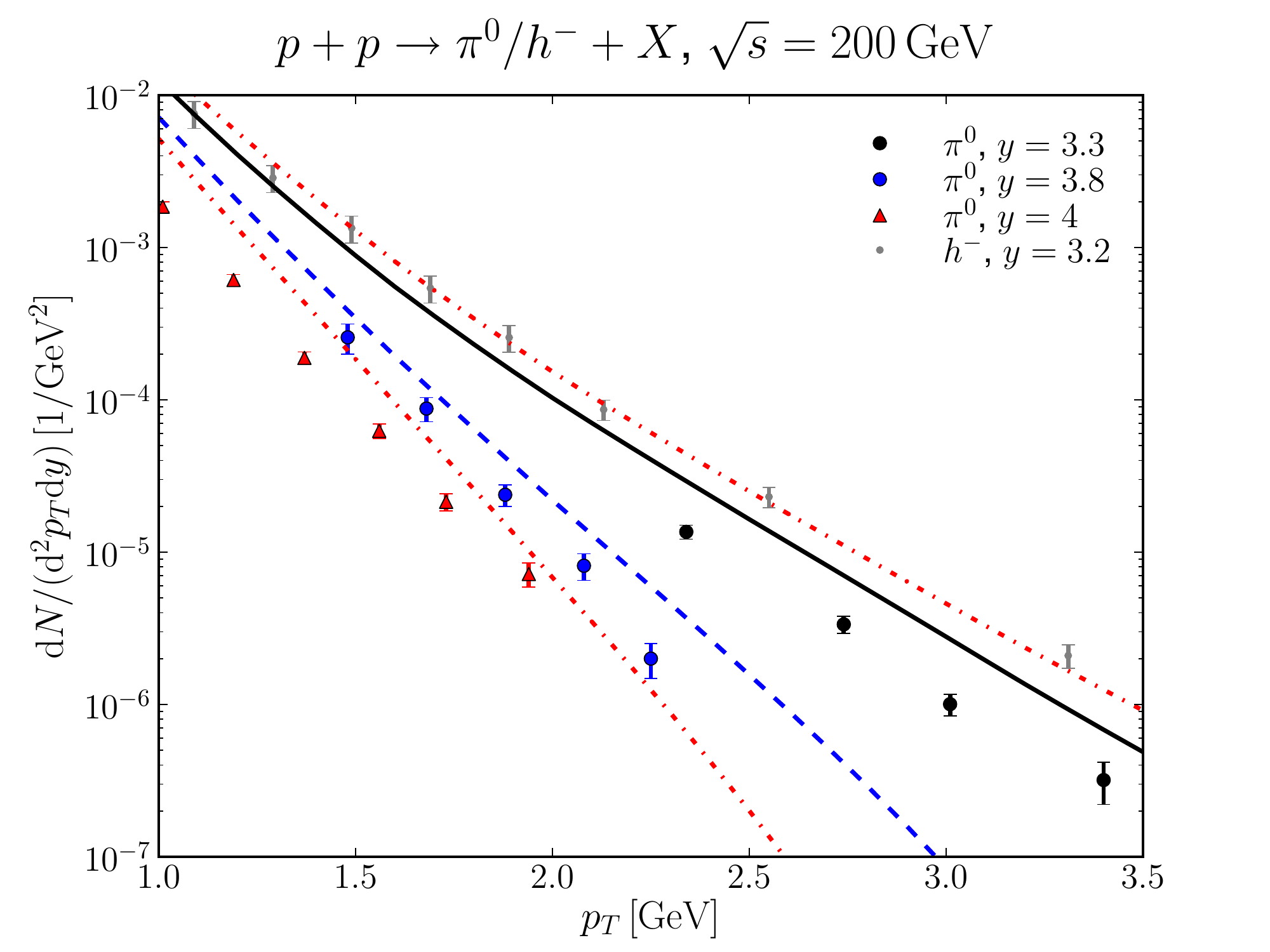} 
\caption{$\pi^0$ production in forward p+p collisions compared with BRAHMS~\cite{Arsene:2004ux} $h^-$ and STAR~\cite{Adams:2006uz} $\pi^0$ data}
\label{fig:pp-yield} 
\end{minipage}
\hspace{0.5cm}
\begin{minipage}[t]{0.48\linewidth}
\centering
\includegraphics[width=1.05\textwidth]{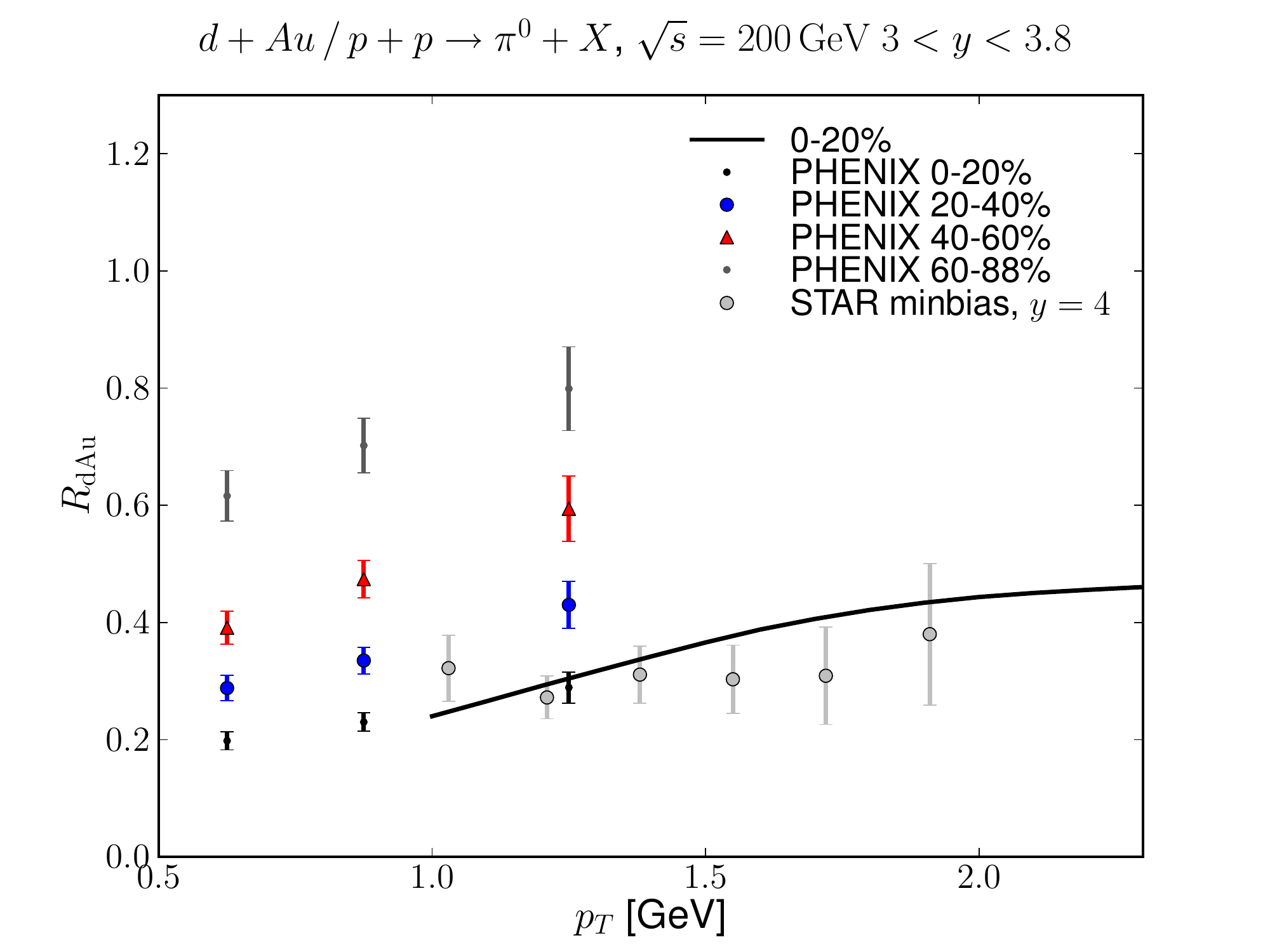} 
\caption{Nuclear suppression factor $R_{\mathrm{dAu}}$ compared with PHENIX~\cite{Adare:2011sc} and STAR~\cite{Adams:2006uz} $\pi^0$ data. Note that the STAR data is at a slightly different rapidity.}
\label{fig:rdau} 
\end{minipage}
\end{figure}

We shall here use the dipole operator obtained from 
solving numerically the BK evolution 
equation using the Balitsky running coupling prescription~\cite{Balitsky:2006wa}.
The BK equation
requires an initial condition, for which we take the McLerran-Venugopalan
model~\cite{McLerran:1994ni}
\begin{equation} \label{eq:bkinitc}
S(r)_{x=x_0} = \exp\left\{
 -\frac{r^2\qs^2}{4}  \ln \left( e + \frac{1}{r^2\lqcd^2}\right)\right\},
\end{equation}
with $Q_s^2=0.2\,\gev^2$ for the proton at $x_0=0.007$ similarly as in \re \cite{Albacete:2010bs}
. Following \re 
\cite{Dumitru:2005gt}
we can then calculate single inclusive hadron production in forward rapidities. The comparison
with STAR $\pi^0$ and BRAHMS charged hadron data is shown in \fig \ref{fig:pp-yield}.
This parametrization overestimates the pion yield by a factor $\sim 3$, but
describes the charged hadron yield correctly. 
We prefer to use the same parameters as \re \cite{Albacete:2010bs} to see clearly the effects of including the 
Gaussian quadrupole. We expect most of the overall normalization error to cancel in the per trigger correlation
and not affect the systematics of the away side peak.

For a nuclear target we use larger initial saturation scale, and we set $\qso^2$ to $0.72\gev^2$ at same $x_0=0.007$ 
which fits the most central PHENIX $R_{\mathrm{dAu}}$ data shown in \fig \ref{fig:rdau}.
However it seems quite impossible to simultaneously describe STAR minimum bias
data.  We consider the normalization issue of the single inclusive baseline as the largest uncertainty in our calculation.

\section{Dihadron correlations}

Consider a large~$x$ quark with momentum $p^+$ from the probe deuteron or proton, 
propagating eikonally through the target nucleus or proton. It can 
radiate a gluon with momentum $k^+=zp^+$ and is left with longitudinal momentum
$q^+ = (1-z)p^+$. In the high energy limit the scattering of both the quark 
and the gluon can be described by the eikonal approximation, where 
they pick up a phase given by a Wilson line in the color field of the
target.  The detailed derivation  of the double inclusive
cross section is performed in \re\cite{Marquet:2007vb} and 
results in the following expression for the $qA\rightarrow qgX$ cross section:

\begin{equation}
\label{eq:dihadron-xs}
\begin{split}
&\frac{\ud \sigma^{qA\to qgX}}{\ud k^+ \ud^2 \kt \ud q^+ \ud^2 \qt } 
= \as \cf \delta(p^+-k^+-q^+) 
\int \frac{\ud^2 \xt}{(2\pi)^2} \frac{\ud^2 \xtp}{(2\pi)^2} 
\frac{\ud^2 \bt}{(2\pi)^2} \frac{\ud^2 \btp}{(2\pi)^2} 
e^{i \kt \cdot(\xtp - \xt)} e^{i\qt \cdot(\btp-\bt)} \\
&\quad\times \sum_{\alpha\beta \lambda} 
\phi_{\alpha\beta}^{\lambda*}(\xtp-\btp) \phi_{\alpha\beta}^\lambda(\xt- \bt) 
\left \{ S^{(4)}(\bt,\xt,\btp,\xtp) - S^{(3)}(\bt,\xt,\ztp) 
 - S^{(3)}(\zt,\xtp,\btp) 
+ S^{(2)}(\zt,\ztp) \right \} ,
\end{split}
\end{equation}
with $\zt = z \xt +(1-z) \bt$ and likewise, $\ztp = z \xtp + (1-z)\btp$. The momenta of the produced gluon and quark are  $\kt$ and $\qt$ respectively.
Likewise, $\xt,\xtp$ should be interpreted as the transverse position
of the gluon and $\bt,\btp$ of the quark, in 
the amplitude and the complex conjugate respectively.
Here $S^{(n)}$ are operators constructed from Wilson lines whose expressions can be found e.g. from \re \cite{Marquet:2007vb}, and $\phi$ is the $q\to qg$ splitting function.
We set the scale at which the parton distribution functions, fragmentation functions and coupling constant are evaluated, 
to be equal to the transverse momenta of the trigger hadron.

\begin{figure}
\begin{minipage}[t]{0.48\linewidth}
\includegraphics[width=\textwidth]{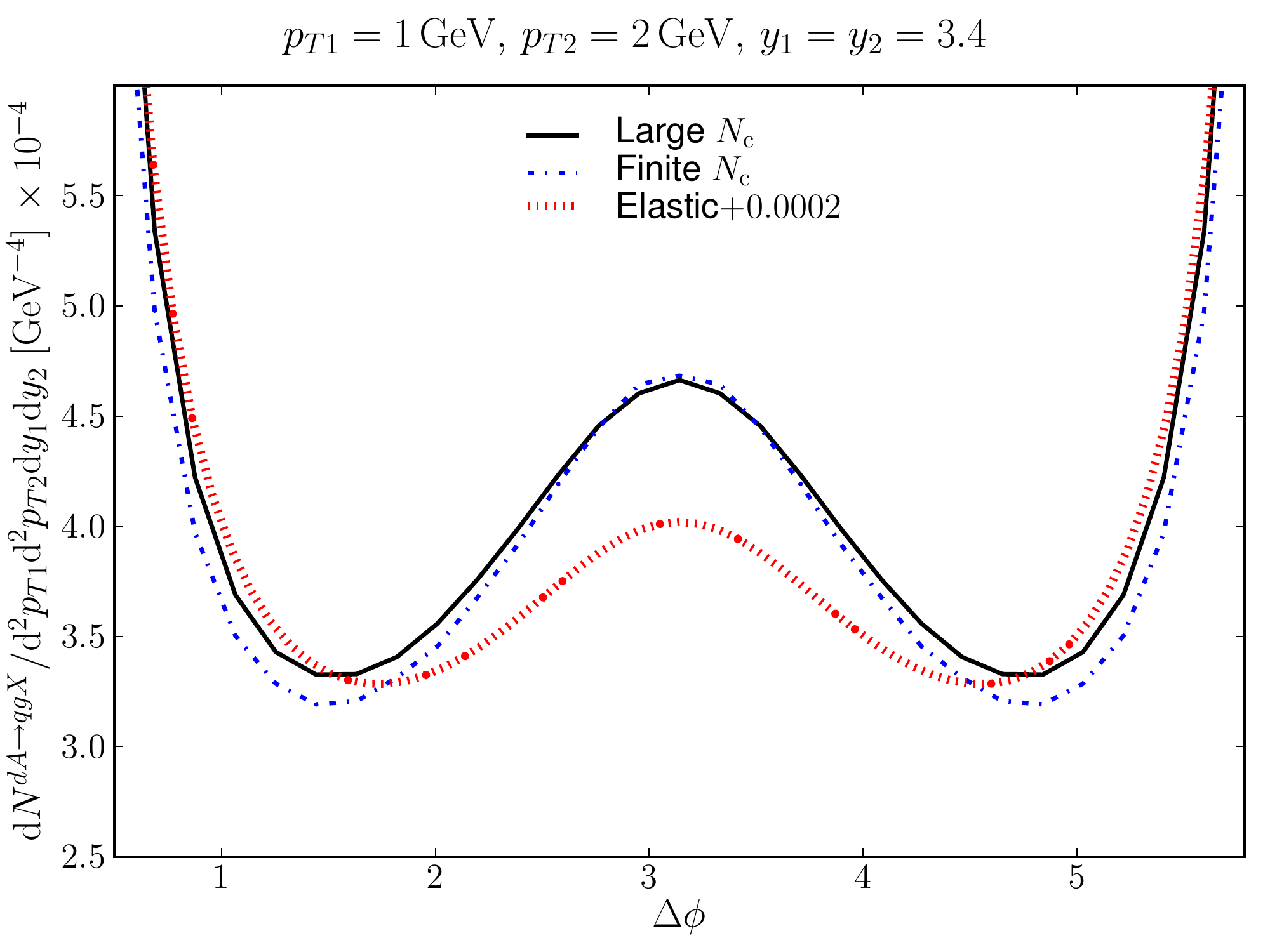} 
\caption{The quark-gluon parton level correlations near forward RHIC kinematics 
Shown are the ``naive large-$\nc$'' approximation (labeled as ``elastic'' in the plot)  used in \cite{Albacete:2010pg} and our Gaussian approximation for the quadrupole and its large-$\nc$ limit. Note that fixed $\Delta \phi$-independent pedestal of $0.0002\,\mathrm{GeV}^{-4}$ is added to the ``elastic'' approximation for the purposes of visualization.}
\label{fig:finite-vs-large}
\end{minipage}
\hspace{0.5cm}
\begin{minipage}[t]{0.48\linewidth}
\includegraphics[width=\textwidth]{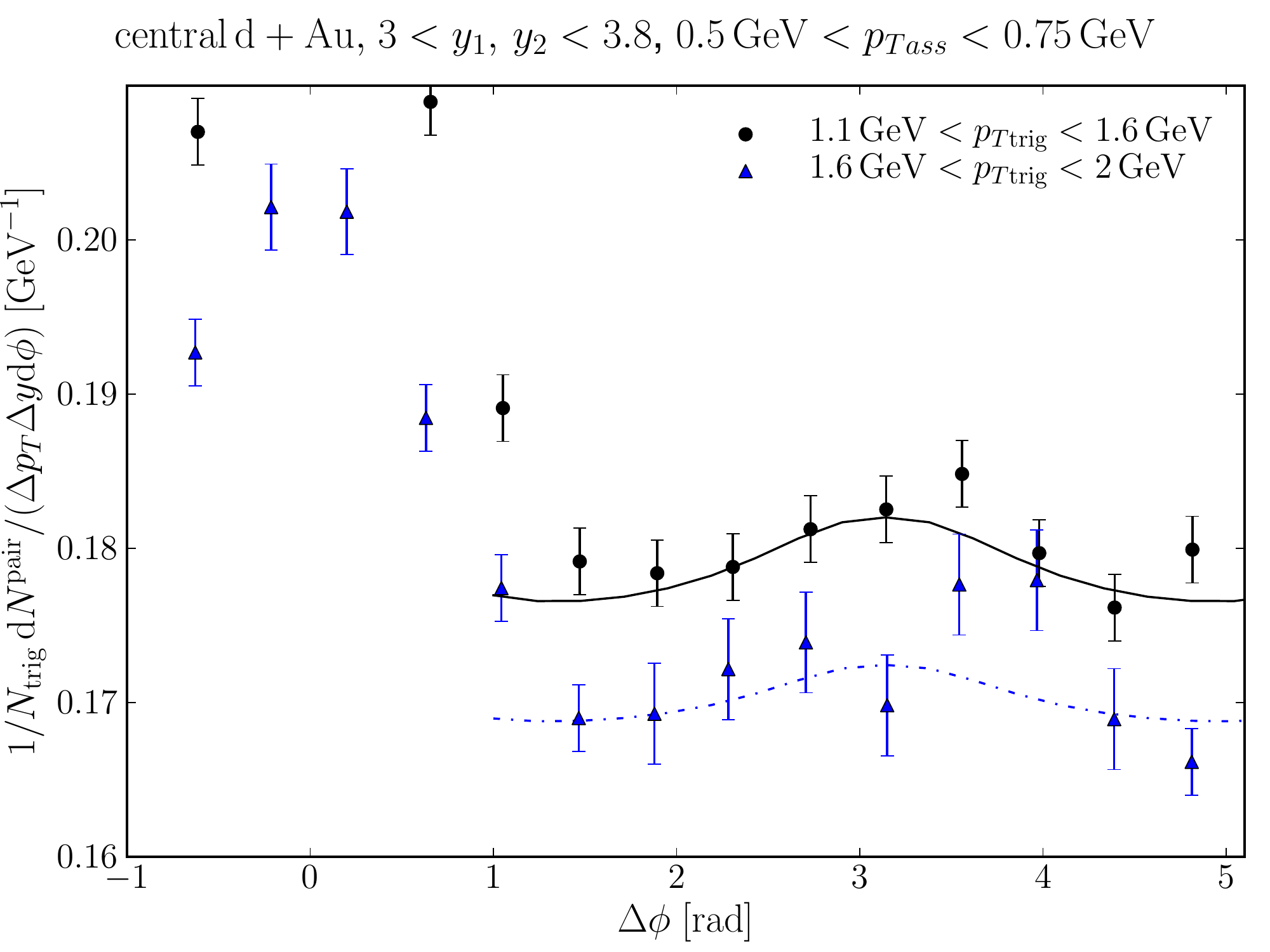} 
\caption{The $\pi^0$ azimuthal angle correlation compared to the PHENIX \cite{Adare:2011sc} dAu result. The $\Delta \phi$ independent pedestal is adjusted to fit the experimental data. The initial saturation scale is $\qso^2=0.72\,\mathrm{GeV}^2$ at $x_0=0.007$ and the large-$\nc$ limit of the Gaussian approximation is used.}
\label{fig:cp}
\end{minipage}
\end{figure}

In \fig \ref{fig:finite-vs-large} we show the parton level dihadron production cross section obtained using different approximations for the quadrupole when evaluating $S^{(4)}$. We notice that the width and especially the height of the away side peak are modified when replacing the naive approximation, used e.g. in \cite{Albacete:2010pg}, by the Gaussian approximation \cite{Dominguez:2011wm}. The comparison with experimental data is shown in \fig \ref{fig:cp}, where we compute the coincidence probability measured by the PHENIX collaboration~\cite{Adare:2011sc}. Once we have adjusted the $\Delta \phi$ independent pedestal contribution we obtain a good description of the $\Delta \phi$ dependent part of the data.

\section{Double parton scattering}
The $\Delta \phi$ independent pedestal in the experimentally measured coincidence probability is mostly due to the double parton scattering contribution where two hadrons are produced independently of each other. In previous works \cite{Albacete:2010pg, Stasto:2011ru} this process has been considered to be completely separated from the correlated dihadron production. 
However, when one uses the Gaussian approximation for the quadrupole function one can show that \eq \eqref{eq:dihadron-xs} actually contains part of this DPS (Double Parton Scattering) contribution. 

The DPS contribution is obtained when gluon is emitted far away from the quark: $|\bt - \btp|\sim |\xt - \xtp| \sim 1/\qs,$ 
$\utt \equiv |\bt-\xt|\gg 1/\qs$. In this limit the only surviving correlation in $S^{(4)}$ 
is the product of an adjoint representation dipole at the location of the
gluon and a fundamental representation one at the location of the quark. We call this kinematical regime the ``DPS'' limit. 
In this limit the Wilson lines of the quark and the gluon are uncorrelated and we have,
using the fact that the expectation values must be color singlets,
\begin{equation}
\label{eq:s4dps}
\begin{split}
&S^{(4)}(\bt,\xt,\btp,\xtp) 
\underset{\textrm{DPS}}{\approx}
S^{(4)}_\textrm{DPS}(\bt,\xt,\btp,\xtp) 
=
\frac{\nc^2}{\nc^2-1}
\left \langle \hat{D}(\bt,\btp)\right \rangle
\left \langle \hat{D}^2(\xt,\xtp) - \frac{1}{\nc^2}\right \rangle,
\end{split}
\end{equation}
where $\hat D$ is a correlator of two Wilson lines (two point function) in the fundamental representation, and the expectation value of $\hat D^2 - 1/\nc^2$
 can be identified
as the two point function in the adjoint representation. 

For massless quarks the four-point function $S^{(4)}$ causes the  
integral \nr{eq:dihadron-xs} to diverge logarithmically in the DPS
limit. Physically this means that the quark emits a
very small 
transverse momentum gluon. The quark and gluon subsequently 
scatter independently off the target. 
The DPS limit corresponds 
to a splitting happening a long time before the interaction with the target.
This logarithmically  divergent contribution must  be regulated by 
confinement scale physics in the wavefunction of the projectile. It is in fact exactly 
the kind of contribution that is represented by double parton scattering,  
discussed in this context e.g. in \re\cite{Strikman:2010bg}.

We add to the dihadron production cross section the
DPS contribution. In order to avoid double counting the logarithmically divergent
DPS part is subtracted from \eq \eqref{eq:dihadron-xs}. This forces us to introduce
an arbitrary (soft) cutoff scale $\sim \lqcd^{-1}$. 
The DPS contribution is then divided into two separate parts. The first one 
corresponds to taking two partons from the same nucleon in the deuteron,
described by a single-nucleon double parton distribution function, which we model by 
implementing a kinematical constraint $x_i + x_j < 1$ following \re \cite{Strikman:2010bg}.
The  second contribution involves taking one parton from the neutron and the other one from 
the proton, which is not bound by the same kinematical constraint. Once we have obtained the
corresponding parton distribution functions, the DPS yield is essentially the single inclusive
yield squared.

As a result we obtain an order-of-magnitude estimate for the $\Delta \phi$ independent pedestal. 
When comparing with PHENIX data, we obtain for the trigger transverse momentum range $1.1\dots 1.6 \gev$
a pedestal $0.11 \gev^{-1}$, whereas the experimental value is $0.176 \gev^{-1}$.
Similarly for the trigger transverse momentum $1.6 \dots 2\gev$ we obtain $0.08 \gev^{-1}$, and
 the experimental value reads $0.163 \gev^{-1}$ \cite{Adare:2011sc}.
 
\section{Conclusions}
We have shown that the previously used ``naive large-$\nc$'' approximation for the 
quadrupole is not accurate, and the width and especially the
height of the back-to-back peak is modified when more accurate Gaussian approximation is used. In addition,
the naive approximation misses an important logarithmically divergent DPS 
contribution, which is included consistently in our work.

We obtain a relatively good estimate for the $\Delta \phi$ independent
pedestal background, and a good description of the $\Delta \phi$ dependent
part of the PHENIX data. We also point out that the ``naive large-$\nc$'' 
approximation clearly underestimates the height of the away side peak.

We thank K. J. Eskola, I. Helenius, R. Paatelainen, 
B. Schenke, M. Strikman and R. Venugopalan
for discussions
and J. Albacete and C. Marquet for helpful comparisons with their results.
H.M. is supported by the Graduate School of Particle and Nuclear Physics.
This work has been supported by the Academy of Finland, projects
141555 and 133005 and by computing resources from CSC - IT Center for Science in 
Espoo, Finland.





\bibliographystyle{h-physrev4mod2}
\bibliography{spires}

\providecommand{\href}[2]{#2}\begin{thebibliography}{10}

\bibitem{Lappi:2010ek}
T.~Lappi,
\newblock \href{http://dx.doi.org/10.1142/S0218301311017302}{Int.J.Mod.Phys.
  {\bf E20}, 1 (2011)}, [\href{http://arXiv.org/abs/1003.1852}{{arXiv:1003.1852
  [hep-ph]}}].

\bibitem{Braidot:2011zj}
E.~Braidot,
\newblock \href{http://arXiv.org/abs/1102.0931}{{arXiv:1102.0931 [nucl-ex]}}.

\bibitem{Adare:2011sc}
PHENIX, A.~Adare {\em et~al.},
\newblock
  \href{http://dx.doi.org/10.1103/PhysRevLett.107.172301}{Phys.Rev.Lett. {\bf
  107}, 172301 (2011)}, [\href{http://arXiv.org/abs/1105.5112}{{arXiv:1105.5112
  [nucl-ex]}}].

\bibitem{Dumitru:2011vk}
A.~Dumitru, J.~Jalilian-Marian, T.~Lappi, B.~Schenke and R.~Venugopalan,
\newblock \href{http://dx.doi.org/10.1016/j.physletb.2011.11.002}{Phys.Lett.
  {\bf B706}, 219 (2011)},
  [\href{http://arXiv.org/abs/1108.4764}{{arXiv:1108.4764 [hep-ph]}}].

\bibitem{Marquet:2007vb}
C.~Marquet,
\newblock \href{http://dx.doi.org/10.1016/j.nuclphysa.2007.09.001}{Nucl. Phys.
  {\bf A796}, 41 (2007)},
  [\href{http://arXiv.org/abs/0708.0231}{{arXiv:0708.0231 [hep-ph]}}].

\bibitem{Albacete:2010pg}
J.~L. Albacete and C.~Marquet,
\newblock
  \href{http://dx.doi.org/10.1103/PhysRevLett.105.162301}{Phys.Rev.Lett. {\bf
  105}, 162301 (2010)}, [\href{http://arXiv.org/abs/1005.4065}{{arXiv:1005.4065
  [hep-ph]}}].

\bibitem{Lappi:2012nh}
T.~Lappi and H.~M{\"a}ntysaari,
\newblock \href{http://arXiv.org/abs/1209.2853}{{arXiv:1209.2853 [hep-ph]}}.

\bibitem{Balitsky:2006wa}
I.~Balitsky,
\newblock \href{http://dx.doi.org/10.1103/PhysRevD.75.014001}{Phys. Rev. {\bf
  D75}, 014001 (2007)},
  [\href{http://arXiv.org/abs/hep-ph/0609105}{{arXiv:hep-ph/0609105}}].

\bibitem{McLerran:1994ni}
L.~D. McLerran and R.~Venugopalan,
\newblock \href{http://dx.doi.org/10.1103/PhysRevD.49.2233}{Phys. Rev. {\bf
  D49}, 2233 (1994)},
  [\href{http://arXiv.org/abs/hep-ph/9309289}{{arXiv:hep-ph/9309289}}].

\bibitem{Albacete:2010bs}
J.~L. Albacete and C.~Marquet,
\newblock \href{http://dx.doi.org/10.1016/j.physletb.2010.02.073}{Phys.Lett.
  {\bf B687}, 174 (2010)},
  [\href{http://arXiv.org/abs/1001.1378}{{arXiv:1001.1378 [hep-ph]}}].

\bibitem{Dumitru:2005gt}
A.~Dumitru, A.~Hayashigaki and J.~Jalilian-Marian,
\newblock \href{http://dx.doi.org/10.1016/j.nuclphysa.2005.11.014}{Nucl. Phys.
  {\bf A765}, 464 (2006)},
  [\href{http://arXiv.org/abs/hep-ph/0506308}{{arXiv:hep-ph/0506308}}].

\bibitem{Arsene:2004ux}
BRAHMS, I.~Arsene {\em et~al.},
\newblock Phys. Rev. Lett. {\bf 93}, 242303 (2004),
  [\href{http://arXiv.org/abs/nucl-ex/0403005}{{arXiv:nucl-ex/0403005}}].

\bibitem{Adams:2006uz}
STAR, J.~Adams {\em et~al.},
\newblock \href{http://dx.doi.org/10.1103/PhysRevLett.97.152302}{Phys. Rev.
  Lett. {\bf 97}, 152302 (2006)},
  [\href{http://arXiv.org/abs/nucl-ex/0602011}{{arXiv:nucl-ex/0602011}}].

\bibitem{Dominguez:2011wm}
F.~Dominguez, C.~Marquet, B.-W. Xiao and F.~Yuan,
\newblock \href{http://dx.doi.org/10.1103/PhysRevD.83.105005}{Phys.Rev. {\bf
  D83}, 105005 (2011)}, [\href{http://arXiv.org/abs/1101.0715}{{arXiv:1101.0715
  [hep-ph]}}].

\bibitem{Stasto:2011ru}
A.~Stasto, B.-W. Xiao and F.~Yuan,
\newblock \href{http://arXiv.org/abs/1109.1817}{{arXiv:1109.1817 [hep-ph]}}.

\bibitem{Strikman:2010bg}
M.~Strikman and W.~Vogelsang,
\newblock \href{http://dx.doi.org/10.1103/PhysRevD.83.034029}{Phys.Rev. {\bf
  D83}, 034029 (2011)}, [\href{http://arXiv.org/abs/1009.6123}{{arXiv:1009.6123
  [hep-ph]}}].

\end{thebibliography}







\end{document}